\documentclass[namedreferences]{kluwer}

\usepackage{epsfig}

\begin{document}
\begin{article}
\begin{opening}

\title{Regular and Chaotic Motion in Globular Clusters}
\author{\surname{Daniel D.~Carpintero}\email{ddc@fcaglp.unlp.edu.ar}}
\author{\surname{Juan C.~Muzzio}\email{jcmuzzio@fcaglp.unlp.edu.ar}}
\author{\surname{Felipe C.~Wachlin}\email{fcw@fcaglp.edu.ar}}
\institute{Facultad de Ciencias Astron\'omicas y Geof\'{\i}sicas - UNLP
and PROFOEG -- CONICET}
\begin{ao}
Observatorio Astron\'omico -- Paseo del Bosque S/N
1900 La Plata -- Buenos Aires
Argentina
\end{ao}
\begin{abstract}
As a first step towards a comprehensive investigation of stellar motions
within globular clusters, we present here the results of a study of
stellar orbits in a mildly triaxial globular cluster that follows a 
circular orbit inside a galaxy. The stellar orbits were classified using the
frequency analysis code of Carpintero and Aguilar and, as a check,
the Liapunov characteristic exponents were also computed in some cases.

The orbit families were obtained using different start spaces. 
Chaotic orbits turn out to be very common and while, as could be expected,
they are particularly abundant in the outer parts of the cluster, they
are still significant in the innermost regions. Their relevance for the
structure of the cluster is discussed.
\end{abstract}
\end{opening}

\section{Introduction}
We tend to think of globular clusters as spherical stationary stellar
systems that are well described by King's or Michie's models (see,
e.g., \opencite{BT87}). Obviously, nobody in his right mind
would search for chaotic motions in such systems, but the truth is
that: a) Globular clusters are not spherical and exhibit different
degrees of ellipticity (see, e.g., \opencite{HR94}); b) Globular clusters are not
isolated systems and the motions of their stars are governed, not only
by the cluster's field, but by the tidal forces of the galaxy
where the cluster belongs as well. Thus, as neither angular momentum 
nor energy has to be conserved, it is very reasonable to expect to
find chaotic motions in the stellar orbits within globular clusters.

The presence of significant chaotic motions would certainly have
important consequences for the structure of the cluster and the
models should take into account this fact. The present work is just
a first step to show that,
even under very simple hypotheses, chaotic orbits turn out to be
very abundant in globular clusters, thus paving the way for future,
more detailed, studies on this subject. 
\section{The model}
We wanted to begin our investigation of chaos in globular clusters
with the simplest possible case, so that:

\begin{itemize}
\item[a)] We adopted a circular orbit for the motion of the
globular cluster around the galaxy.

\item[b)] We assumed that the cluster is deformed by the effect of the
tidal forces only.

\item[c)] We neglected the effects of stellar encounters within the cluster.
\end{itemize}

The adoption of more realistic conditions should increase
chaos because: a) With elongated cluster orbits we lose
Jacobi's integral; b) More triaxial potentials might enhance
chaoticity; c) Impulsive forces will contribute to chaos.

The galaxy was represented by a spherically symmetrical
logarithmic potential and the
globular cluster with a modified Satoh distribution, whose
potential is:

\begin{equation}
\Phi_S(x,y,z)=-{GM\over \sqrt{x^2+y^2+z^2+g(
g+2\sqrt{y^2+(z/b)^2+h^2})}}.
\end{equation}

Here the origin of coordinates lies at the center of the globular
cluster, the $x$ axis points in the direction opposite to the galactic 
center, the $y$ axis in the direction of motion of the cluster
around the galaxy and the $z$ axis perpendicular to the orbital 
plane.
Notice that we have interchanged $x^2 + y^2$ with $z^2$, in order
to obtain a prolate (rather than Satoh's oblate) system, and
that we have also divided $z$ by a parameter $b$, in order to get
a triaxial system (tidal deformation yields the shortest axis 
perpendicular to the orbital plane).

\begin{figure}[t]
\centerline{\epsfbox[94 432 419 728]{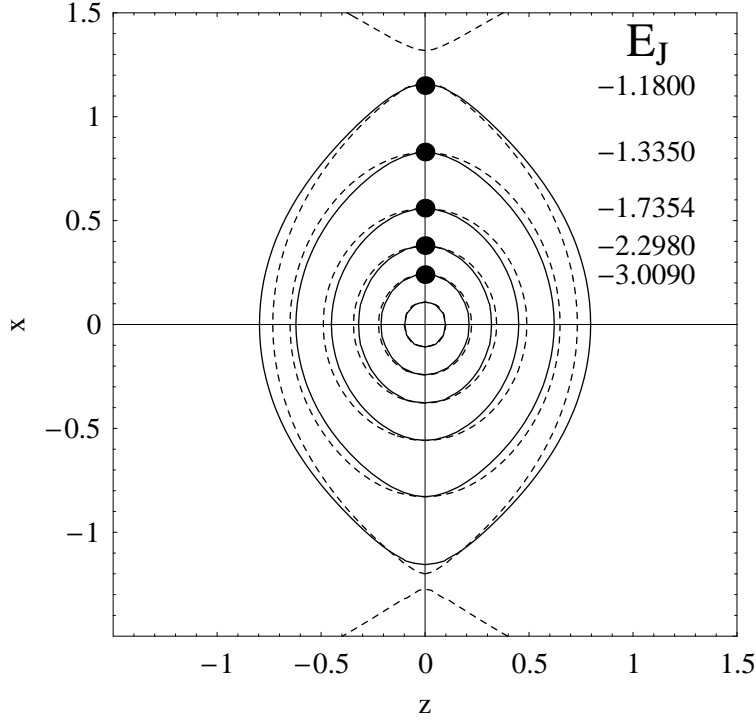}}
\caption{Isodensity curves for the modified Satoh potential
(full lines) in the $x$--$z$ plane. The equipotentials that
result from adding the centrifugal term and the galactic potential 
are shown as dashed lines.}
\label{isodensitas}
\end{figure}

The main advantage of this election is that isodensity 
surfaces increase their ellipticity as we move outwards,
just as it should happen with a system that is tidally
deformed, as shown by the full lines in Figure \ref{isodensitas}. We
have also included in the figure (dashed lines) the
effective equipotential curves (i.e., those that result
from adding the centrifugal term and the galactic potential to the modified Satoh
potential). 

The equations of motion are:

\begin{equation}
\ddot{x}=-{GMx\over S^3}-\omega^2 R^2 {R+x\over (R+x)^2
+y^2+z^2}+\omega^2 (R+x)+2\omega \dot y;
\end{equation}
\begin{equation}
\ddot{y}=-{GMy(1+g/T)\over S^3}-\omega^2 R^2 {y\over (R+x)^2
+y^2+z^2}+\omega^2 y-2\omega \dot x;
\end{equation}
\begin{equation}
\ddot{z}=-{GMz(1+g/(b^2 T))\over S^3}
-\omega^2 R^2 {z\over (R+x)^2+y^2+z^2},
\end{equation}
where
\begin{equation}
S=\sqrt{x^2+y^2+z^2+g(g+2T)},\ \ 
T=\sqrt{y^2+(z/b)^2+h^2},
\end{equation}
and $M$ is the mass of the globular cluster, 
$R$ is the radius of its orbit, and
$\omega$ is its angular velocity.
The Jacobi integral is:
\begin{equation}
E_J={1\over 2} (\dot{x}^2+\dot{y}^2+\dot{z}^2)-
    {1\over 2} \omega^2 \left[ (R+x)^2 + y^2 \right] 
    + \Phi(x,y,z),
\end{equation}
where $\Phi$ is the sum of the potential of the globular cluster, $\Phi_S$,
and that of the galaxy:
\begin{equation}
\Phi_G(x,y,z)=\frac{1}{2} \omega^2 R^2 \ln \left[ (R+x)^2+y^2+z^2 \right].
\end{equation}

We adopted the following values:
$ b = 0.8,\, h = 0.5,\, g = 0.05,\, R = 100,\, \omega = 0.5$,
which result in a tidal radius
$ r_t=x_t = 1.24$,
and the half--mass radius is $r_h = 0.28$.

If, for example, we choose $R = 10$ kpc and the mass of the
galaxy within that radius as $M_g = 1.25\times 10^{11} M_\odot$, then we 
have a tidal radius of about $r_t = 120$ pc and a cluster mass
of $M = 5\times 10^5 M_\odot$, that is, reasonable values for
a globular cluster.

\section{Orbital Analysis}

\subsection{Liapunov characteristic exponents}

D.~Pfenniger kindly let us use his {\tt LIAMAG} routine that
computes the six Liapunov exponents following Benettin's
method.

We are not so much interested on whether a specific stellar orbit is chaotic
or not, as in having statistical information on a large number
of orbits. Therefore, we followed an approach similar to that
of \inlinecite{MF96}: 1) We integrated the orbits
for about 10,000 orbital periods (rather than 100, as they
did); 2) We used as estimator the sum of the three non--negative 
Liapunov exponents (also called Kolmogorov entropy);
3) We dubbed, rather arbitrarily, chaotic those orbits where:

\begin{equation}
\ln (s_1 + s_2 + s_3) > -5,
\end{equation}
where the $s_i (i = 1, 2, 3)$ are the positive estimates of the
Liapunov exponents after 10,000 orbital periods.

We used this method just as a check, because:
1) It only allows one to decide between regular and chaotic
orbits, providing no further information on the kind of orbit
one has; 2) It is very slow (about one day of computing on a
Pentium Pro, 200 MHz, personal computer for 150 orbits).

\subsection{Frequency Analysis}
This technique was introduced by \citeauthor{BS82} (\citeyear{BS82},
\citeyear{BS84}) and extended, in a different form, by \inlinecite{L93}.
\inlinecite{CA98} refined the original method
and prepared a {\tt FORTRAN} code that allows one to automatically
classify large numbers of orbits. 

The basis of the method is that regular orbits move on a
torus--like manifold and are quasi--periodic. Fourier 
spectra of the time series of the coordinates of a regular
orbit consist of discrete lines whose frequencies are integer 
linear combinations of the frequencies of the angle variables.
Thus, from the Fourier spectra one can classify the regular 
orbits. Besides, chaotic orbits yield continuous spectra and
can be recognized too (see \opencite{CA98}, for
examples of different orbits).
The main limitation of the method is the difficulty to
recognize whether finite precision numbers have a rational
quotient.

\subsection{Initial Conditions}
We prepared sets of initial conditions for several values of the 
Jacobi integral ($-3.0090$, $-1.7354$, $-1.3350$ and $-1.1800$). For each
value of the integral, we selected four sets: 1) Zero initial
velocity; 2) $x$--$y$ plane and $\dot z$ initial velocity; 3) $x$--$z$
plane and $\dot y$ initial velocity; 4) $y$--$z$ plane and $\dot x$ initial
velocity.

\inlinecite{S93} proposed the use of the initial
conditions 1 and 3 for non--rotating potentials. As we have a
rotating potential, we preferred to add also initial conditions
2 and 4.
We expect to have sampled the whole phase space with these
sets, at the price of some possible overlap,
as we will see later on.

\section{Results}

\begin{figure}[t]
\centerline{\epsfig{file=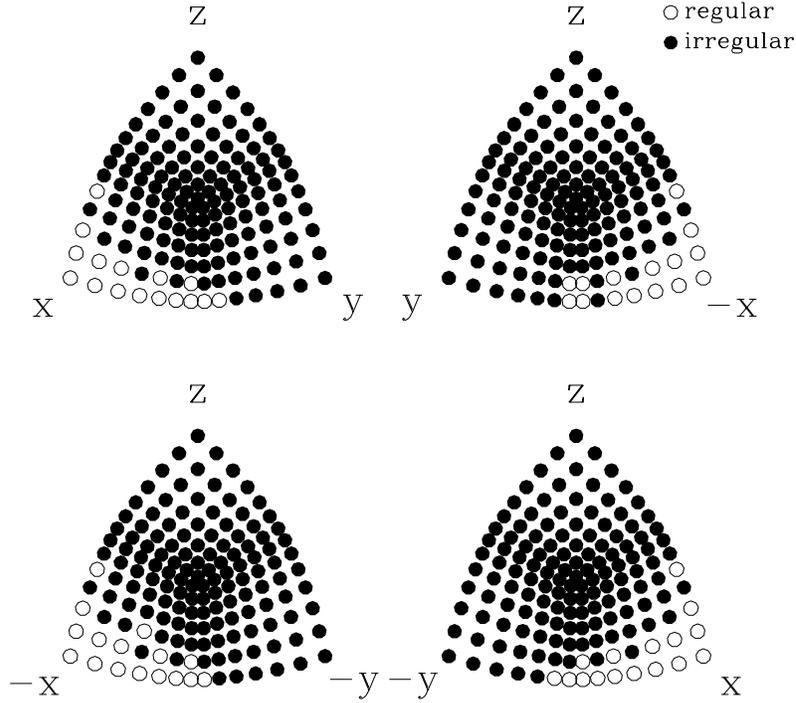,width=12cm}}
\caption{Zero velocity start space for $E_J=-1.335$.
The regular or irregular character of the stellar orbits was
decided from the Liapunov exponents analysis.}
\label{vzerol}
\end{figure}

\begin{figure}[t]
\centerline{\epsfig{file=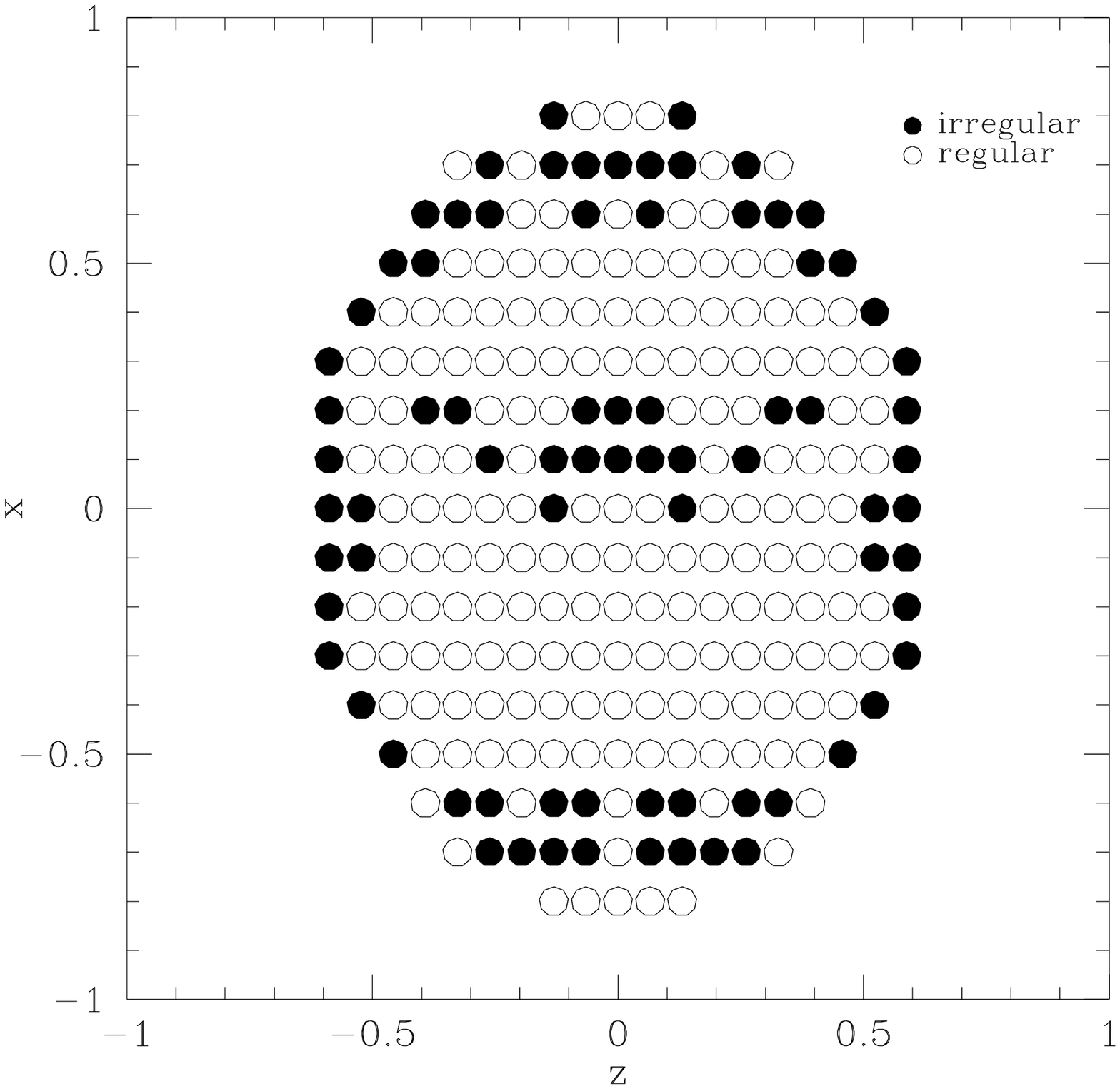,width=10cm}}
\caption{$x$--$z$ start space for $E_J = -1.335$. 
The regular or irregular character of the stellar orbits was
decided from the Liapunov exponents analysis.}
\label{xzl}
\end{figure}

\begin{figure}[t]
\centerline{\epsfig{file=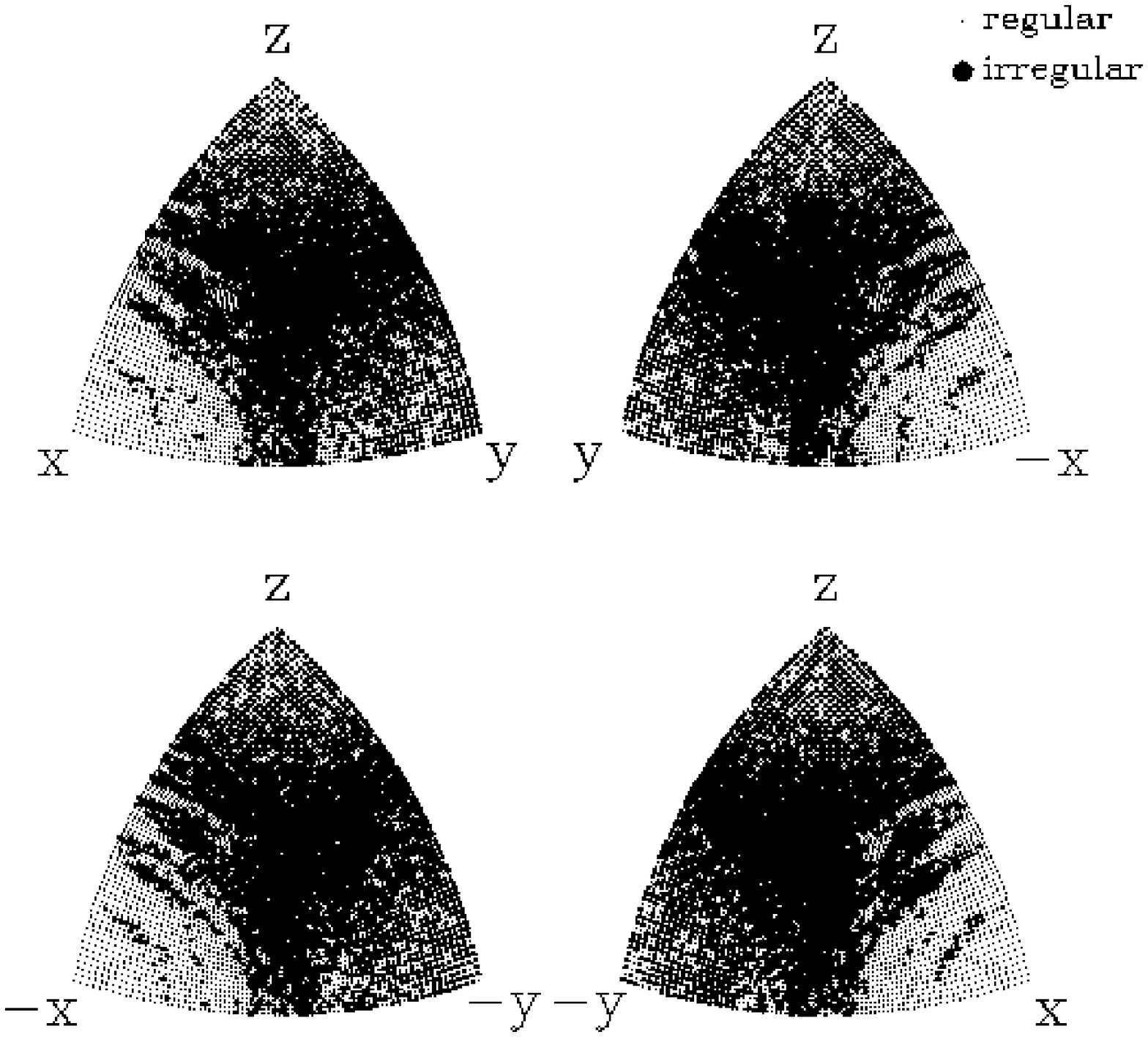,width=12cm}}
\caption{Zero velocity start space for $E_J=-1.335$.
The regular or irregular character of the stellar orbits was
decided from the frequency analysis.}
\label{vzerof}
\end{figure}

\begin{figure}[t]
\centerline{\epsfig{file=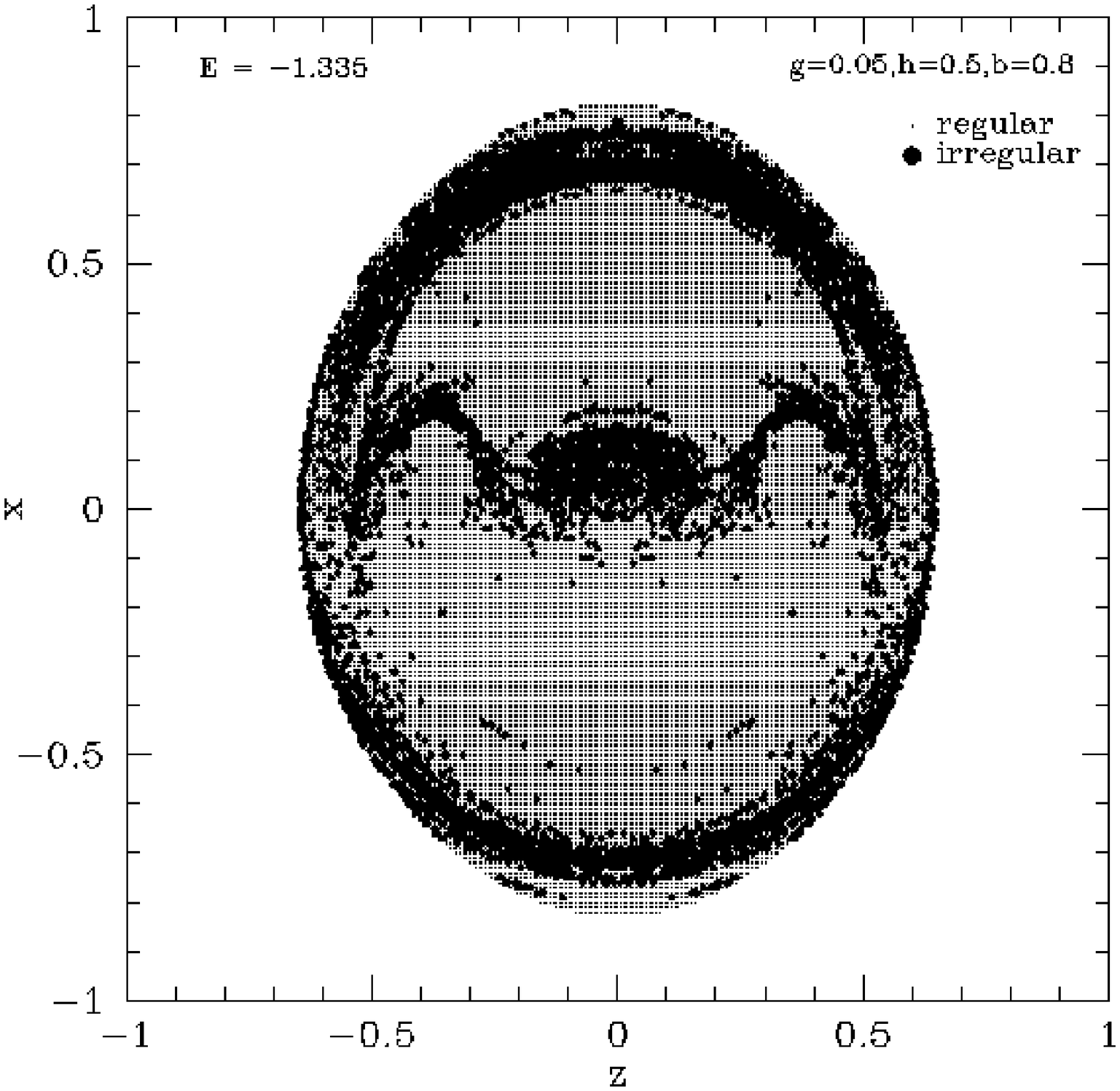,width=12cm}}
\caption{$x$--$z$ start space for $E_J = -1.335$. 
The regular or irregular character of the stellar orbits was
decided from the frequency analysis.}
\label{xzf}
\end{figure}

For every set of initial conditions we prepared colour plots
showing, with different colours, the different types of stellar orbits
that result from those initial conditions. Black and white
examples are given in Figures \ref{vzerol} through \ref{xzf}, which correspond
to a value of the Jacobi integral of $E_J = -1.335$.
Figures \ref{vzerol} and \ref{xzl} are results obtained
from the Liapunov exponents analysis: Figure \ref{vzerol} corresponds to
zero velocity initial conditions and Figure \ref{xzl} to initial
conditions in the $x$--$z$ plane. Figures \ref{vzerof} and \ref{xzf} are results
from the frequency analysis and correspond, respectively, to
the same initial conditions of Figures \ref{vzerol} and \ref{xzl}. Notice that
the density of points is two orders of magnitude larger for
Figures \ref{vzerof} and \ref{xzf}, resulting in much better definition, thanks
to the short computing times needed for the frequency analysis.
There is a generally good agreement between the results of both
methods, but there are a couple of caveats. First, a small
fraction of the orbits (less than 10\%) could not be classified
with the frequency analysis code; from our experience with that
code, we know that most of those orbits turn out to be chaotic
on a more detailed analysis, so that we counted them as such.
Second, the Liapunov exponents tend to give somewhat larger
fractions (by about 10\%) of chaotic orbits. The most likely
explanation for this discrepancy comes from the very different 
integration times: between 100 and 200 periods for the frequency
analysis and 10,000 periods for the Liapunov one; as a result,
orbits that behave regularly most of the time, although they are
truly chaotic, have a much larger chance of getting detected in
the second case.

We noticed that the $x$--$z$ and $y$--$z$ initial conditions gave the
same fractions for the different kinds of orbits, so that we
suspect that with those sets of initial conditions we are
sampling esentially the same parameter space. Therefore, we
combined the results of both start spaces together in
what follows.

\begin{figure}[t]
\centerline{\epsfxsize=12cm\epsfbox[18 459 572 682]{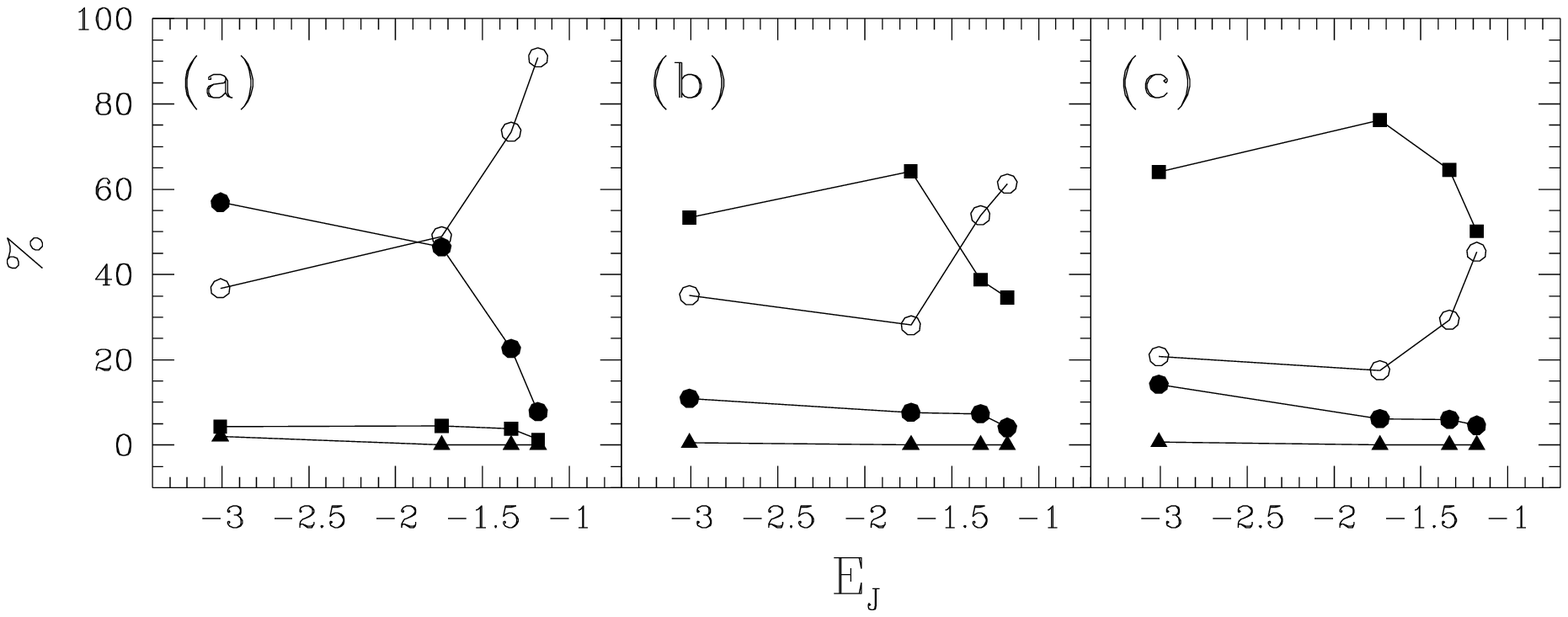}}
\caption{Fractions of the different kinds of stellar
orbits , for $(a)$ the zero initial
velocity, $(b)$ $x$--$y$ and $(c)$ $x$--$z$ plus $y$--$z$ start spaces. 
Filled circles: box orbits; filled triangles: long--axis
tubes; filled squares: short--axis tubes; open circles: chaotic
orbits.}
\label{fracciones}
\end{figure}

Figures \ref{fracciones}$a$, $b$ and $c$ give the fractions of the different
kinds of stellar orbits,
as a function of the Jacobi integral, for the zero initial
velocity, $x$--$y$ and $x$--$z$ plus $y$--$z$ start spaces, respectively.
Boxes and chaotic orbits clearly dominate for zero initial
velocity conditions, while small--axis tubes are the most
abundant orbits in the other cases. Long--axis tubes are
almost non--existent. As expected, chaotic orbits 
predominate for low absolute values of the Jacobi integral,
i.e., mainly in the outermost parts of the globular cluster.
Nevertheless, chaos is still significant for $E_J = -3.009$,
with 37\% of the orbits with zero initial velocity,
35\% of those on the $x$--$y$ start space and 21\% of those on the $x$--$z$ and
$y$--$z$ start spaces.
These results are
particularly important because the zero velocity 
$E_J = -3.009$ surface encloses about 50\% of the total
mass of the globular cluster, so that chaos is present
well inside the cluster and, moreover, seems to dominate
in its outer half.

\section{Conclusions}
From a methodological point of view, we see that the results
of frequency analysis are in generally good agreement with
those from the Liapunov exponents. The advantages of frequency analysis
over Liapunov exponents are that the former needs much less computing time 
and, in addition to decide between regular
and chaotic motion, it also allows the classification
of the regular orbits.

Stellar orbits within globular clusters are highly chaotic. 
For the stars that (barely) reach the tidal limiting surface,
the fraction of chaotic orbits may lie somewhere in between
50\% and 90\%. Nevertheless, it is even more surprising
that the innermost parts of the cluster are also affected,
and that as much as about 30\% of the orbits that reach
the half mass limiting surface might be chaotic.

Moreover, the Liapunov times we obtained for $E_J = -1.335$ are
surprisingly short: they tend to crowd near 10 to 30 time
units, while orbital periods are in the range between 1.4
and 7.5 time units. Not only is this short in terms of orbital
periods but also in terms of cluster age: for the reasonable
choice of units mentioned above, the cluster age would be
about 50 to 100 time units, that is, longer than the
Liapunov times. Evolution should thus be very fast, at least
in the outermost parts of the cluster.

Long--axis tube orbits are very rare, even at the innermost
parts of the cluster, while short--axis tubes are the most
common orbits for non--zero initial velocity conditions.
Box orbits only seem to dominate in the innermost parts, for
zero velocity initial conditions. The scarcity of box orbits
in the outermost, and most elongated, parts of the cluster
may pose some problems to build self--consistent models.
Such models should probably rely on the more abundant chaotic 
orbits but that might, in turn, complicate the building of
stationary models, particularly considering the short Liapunov
times detected here.

\begin{acknowledgements}
We are very grateful to J.~A.~N\'u\~nez for very useful 
suggestions and advice, and to R.~E.~Mart\'{\i}nez, H.~R.~Viturro, 
E.~Su\'arez, M.~C.~Fanjul de Correbo and S.~D.~Abal
de Rocha for their technical assistance. This investigation
was supported with grants from the Universidad Nacional de
La Plata and the Consejo Nacional de Investigaciones
Cient\'{\i}ficas y T\'ecnicas de la Rep\'ublica Argentina.
\end{acknowledgements}

\end{article}
\end{document}